\documentclass[prd,twocolumn,showpacs,superscriptaddress,nofootinbib,floatfix,showkeys,10pt]{revtex4-2}

\usepackage{amsmath,amssymb,amsfonts}
\usepackage{bm}
\usepackage{braket}

\usepackage{array}
\usepackage{multirow}
\usepackage{float}

\usepackage{graphicx}
\usepackage[usenames,dvipsnames]{color}
\usepackage{hyperref}
\usepackage{subcaption}

\usepackage{tikz}
\usetikzlibrary{arrows.meta, calc, positioning}

\usepackage{slashed}
\usepackage{orcidlink}
\usepackage{appendix}
\usepackage{dashrule}

\allowdisplaybreaks[1]

\definecolor{nicered}{rgb}{0.7,0.1,0.1}
\definecolor{nicegreen}{rgb}{0.1,0.5,0.1}
\definecolor{emph}{rgb}{1,0,0}
\definecolor{doub}{rgb}{0.7,0.2,1.0}
\definecolor{navyblue}{RGB}{0, 110, 184}

\hypersetup{colorlinks,citecolor=nicegreen,linkcolor=nicered,urlcolor=navyblue}

\graphicspath{{images/}}

\newcommand{\qbar}{\bar{q}}

\newcommand{\bbar}{\bar{b}}
\newcommand{\cbar}{\bar{c}}
\newcommand{\sbar}{\bar{s}}

\begin{document}
 
\title{Landscape of pentaquark bound states with neural-network Variational Monte Carlo}

\author{Jing-Zhe Song\,\orcidlink{0009-0004-5608-065X}}
\affiliation{School of Physics, Peking University, Beijing 100871, China}

\author{Wei-Lin Wu\,\orcidlink{0009-0009-3480-8810}}\email{wlwu@pku.edu.cn}
\affiliation{School of Physics, Peking University, Beijing 100871, China}

\author{Shi-Lin Zhu\,\orcidlink{0000-0002-4055-6906}}\email{zhusl@pku.edu.cn}
\affiliation{School of Physics and Center of High Energy Physics,
Peking University, Beijing 100871, China}

\begin{abstract}
We present a systematic study of the ground states of pentaquark systems
using the neural network variational Monte Carlo method within the constituent quark potential model. We investigate all manifestly exotic pentaquark systems with negative parity, which cannot mix with the conventional baryons through the creation and annihilation of light or strange quark pair, including $QQQQ\bar{Q}$, $QQQQ\qbar$, $QQQQ\sbar$, $QQQs\bar{Q}$, $QQQq\bar{Q}$, $QQQs\qbar$, $QQQq\sbar$, $QQss\bar{Q}$, $QQsq\bar{Q}$, $QQqq\bar{Q}$, $QQss\qbar$, $QQqq\sbar$, $Qsss\bar{Q}$, $Qssq\bar{Q}$, $Qsqq\bar{Q}$, $Qqqq\bar{Q}$, $Qsss\qbar$, $Qqqq\sbar$, $ssss\bar{Q}$, $sssq\bar{Q}$, $ssqq\bar{Q}$, $sqqq\bar{Q}$, $qqqq\bar{Q}$, $ssss\qbar$, $qqqq\sbar$,  with $Q=b/c$ and $q=u/d$. We exhaust 387  color, spin, and isospin configurations and identify 21 pentaquark bound state candidates with binding energies ranging from $2$ to $15$~MeV. A striking observation is that all of the bound states are shallowly bound molecular states, and a deeply bound pentaquark state does not exist. The $QQqq\bar{Q}$, $ssqq\bar{Q}$, $Qsqq\bar{Q}$ and $Qqqq\bar{Q}$ bound states are composed of a heavy vector meson and a baryon, while the $QQqq\bar{s}$ and $Qqqq\bar{s}$ bound states are composed of a $K$ meson and a heavy baryon. As a cross-validation, the ground-state energies of the unbound systems converge to the baryon--meson thresholds from above, which are consistent with the S-wave scattering states and  provide a reliable baseline for the bound-state criterion. Especially, we reproduce the lowest scattering states $\eta_c N$ and $\eta_c \Lambda$ in the $qqqc\cbar$ and $qqsc\cbar$ channels respectively, where the hidden-charm pentaquark resonances were observed experimentally. We also find that all fully heavy pentaquark states are unbound. Our results provide valuable guidance for future pentaquark searches at LHCb, Belle~II, and other experiments.
\end{abstract}

\maketitle

\section{Introduction}
\label{sec:introduction}
Hadron spectroscopy is one of the primary tools for understanding the nonperturbative nature of Quantum Chromodynamics (QCD).
In 1964, Gell-Mann~\cite{gellmann1964} and Zweig~\cite{zweig1964} independently proposed the quark model,
using three flavors of quarks ($u$, $d$, $s$) to classify
the mesons and baryons known at the time.
However, the color SU(3) gauge symmetry of QCD in principle allows richer color-singlet combinations beyond the conventional mesons and baryons, such as tetraquark 
$qq\bar{q}\bar{q}$, pentaquark $qqqq\bar{q}$, hybrid $q\bar{q}g$,
and glueball.
Gell-Mann and Zweig already noted this possibility
when they first proposed the quark model.
The exploration of these exotic hadron states---both experimental searches and theoretical predictions---
is one of the most active areas in hadron physics.
It tests our understanding of the QCD confinement mechanism and may reveal new dynamics beyond the naive quark model.

On the experimental side, since the Belle Collaboration discovered the first  exotic hadron candidate $X(3872)$~\cite{belle2003} in 2003,
exotic hadron spectroscopy has developed rapidly.
The mass of $X(3872)$ lies extremely close to the $D^0 \bar{D}^{*0}$ threshold,
and its prominent isospin-violating decay modes cannot be explained easily by the conventional charmonium picture~\cite{belle2003,besiii2019_x3872}.
In the tetraquark sector, the Belle Collaboration reported two charged bottomonium-like states $Z_b(10610)$ and $Z_b(10650)$ in 2011 \cite{Belle:2011aa}. Subsequently, BESIII and Belle independently observed the charged charmonium-like state $Z_c(3900)^\pm$~\cite{besiii2013,belle2013} in 2013, which is the first confirmed exotic state containing at least four quarks.
In 2022, LHCb discovered the doubly-charm tetraquark state $T_{cc}^+$~\cite{lhcb2022a,lhcb2022b},
which lies only about $360$~keV below the $D^0 D^{*+}$ threshold, exhibiting the characteristics of an extremely shallow hadronic molecular state.
In 2020, LHCb observed the fully-charm tetraquark structure $X(6900)$~\cite{lhcb2020},
later confirmed by CMS~\cite{cms2024} and ATLAS~\cite{atlas2023}.
CMS recently reported the first spin-parity measurement of the fully-charm tetraquarks~\cite{cms2025}.

In the pentaquark sector, LHCb discovered the hidden-charm pentaquark resonance states $P_{c\cbar}(4380)^+$ and $P_{c\cbar}(4450)^+$
in $\Lambda_b^0 \to J/\psi K^- p$ decays in 2015~\cite{lhcb2015}.
In 2019, LHCb reanalyzed the data with a larger sample, confirming that $P_{c\cbar}(4450)^+$ actually consists of
two overlapping narrow resonances $P_{c\cbar}(4440)^+$ and $P_{c\cbar}(4457)^+$, and discovered a new narrow state $P_{c\cbar}(4312)^+$~\cite{lhcb2019}.
The masses of these three narrow $P_{c\cbar}$ states all lie near the $\Sigma_c^{(*)} \bar{D}^{(*)}$ thresholds, exhibiting typical characteristics of hadronic molecules. The LHCb Collaboration subsequently reported evidence for a strange hidden-charm pentaquark candidate, $P_{c\cbar s}(4459)^0$, in the $J/\psi\Lambda$ invariant-mass spectrum of $\Xi_b^- \to J/\psi\Lambda K^-$ decays, and observed another state, $P_{c\cbar s}(4338)^0$, in $B^- \to J/\psi\Lambda\bar{p}$ decays~\cite{lhcb_pcs4459,lhcb_pcs4338}. Interestingly, hidden-charm molecular configurations corresponding to the observed $P_{c\cbar}$ and $P_{c\cbar s}$ states have been theoretically anticipated before the experimental observations~\cite{Wu:2010vk,Wu:2010jy,Wang:2011rga,Yang:2011wz}. According to the latest Review of Particle Physics~\cite{pdg2026}, the number of exotic hadron candidates with established or strong experimental evidence has reached several dozen and continues to grow.
The current experimental progress places a dual demand on theoretical studies: elucidating the nature of the observed states, and predicting new states to guide experimental searches. Further discussions of multiquark spectroscopy and the structures of exotic hadrons can be found in comprehensive reviews~\cite{hidden_charm_review,hosaka2016,lebed2017,guo2018,liu2019_review,xyz_review,chen2023_review,meng_review2023,Liu:2024uxn}.

On the theoretical side,  lattice QCD provides a first-principles framework for hadron spectroscopy, but calculations of multiquark systems remain computationally challenging and demanding. Effective field theories offer another systematic approach at low energies, whose predictive power depends on low-energy constants that often need to be constrained by experimental inputs. Phenomenological models therefore serve as an important complementary tool for studying the hadron spectrum. In particular, constituent quark models provide an intuitive and effective framework for investigating multiquark states~\cite{yang2019_doublyheavy,ma2024_dmc,meng2023,chen2024_unified,wu2024_benchmark,meng2025_fullheavy,hiyama2025_pcs}, and are well suited for systematic investigations over a wide range of flavor, spin, and color configurations. Within constituent quark models, quantitative predictions for multiquark states require an accurate solution of the corresponding few-body Schr\"odinger equation. This becomes increasingly challenging as the number of particles grows. Conventional few-body methods have achieved considerable success, but each encounters practical limitations for pentaquark systems. In the Gaussian Expansion Method~\cite{gaussian}, the number of basis size grows exponentially with the particle number. Diffusion Monte Carlo~\cite{foulkes2001} relies on approximations to control the sign problem. Variational Monte Carlo (VMC), on the other hand, is flexible and scalable, but its accuracy is ultimately determined by the expressive power of the trial wave function. These considerations motivate the use of neural networks as flexible variational representations of multiquark wave functions.

In 2017, Carleo and Troyer~\cite{carleo2017} introduced Neural Network Quantum States (NQS), using artificial neural networks to represent quantum many-body wave functions within the variational Monte Carlo framework. Since then, neural-network VMC (NN-VMC) has found broad applications in quantum chemistry~\cite{ferminet,paulinet}, nuclear physics~\cite{adams2021,yang2023,yang2025,wu2026}, and other quantum many-body problems~\cite{ml_physics}. Recently, we developed the DeepQuark framework~\cite{deepquark},
extending NN-VMC to address the unique challenges in multiquark systems. DeepQuark achieves accuracy comparable to other state-of-the-art few-body approaches in baryon and tetraquark systems, and demonstrate its potential for applications to larger multiquark systems.

In this paper, we extend the DeepQuark framework to systematically study pentaquark systems in the constituent quark potential model. We perform calculations for 387 pentaquark configurations with spin $\frac{1}{2},\frac{3}{2},\frac{5}{2}$ and negative parity, including
\begin{itemize}
    \item $QQQQ\bar{Q},$
    \item $QQQQ\qbar,QQQQ\sbar, QQQs\bar{Q}, QQQq\bar{Q}, $
    \item $QQQs\qbar,QQQq\sbar, QQss\bar{Q},QQsq\bar{Q}, QQqq\bar{Q}, $
    \item $QQss\qbar, QQqq\sbar,Qsss\bar{Q},Qssq\bar{Q},Qsqq\bar{Q}, Qqqq\bar{Q},$
    \item $ Qsss\qbar, Qqqq\sbar,ssss\bar{Q},sssq\bar{Q},ssqq\bar{Q},sqqq\bar{Q},qqqq\bar{Q},$
    \item $ssss\qbar, qqqq\sbar.$
\end{itemize}
Here $Q=b/c$ and $q=u/d$. We focus on systems that are manifestly exotic in the sense that they cannot mix with conventional baryons through the creation and annihilation of $q\qbar$ or $s\sbar$ pair. We compare the pentaquark ground-state energies with the corresponding lowest baryon--meson threshold and distinguish bound states from scattering states. The inner structures of the bound states are further studied by examining their color configurations and inter-quark root-mean-square radii. We identify 21 shallow pentaquark bound states with molecular configurations and discuss their possible decay channels.


\section{Theoretical Framework}
\label{sec:theory}

\subsection{Quark Potential Model}

We adopt the AL1 potential model proposed by Semay and Silvestre-Brac~\cite{silvestre1994,silvestre1996},
a nonrelativistic constituent quark model
whose parameters were determined by fitting the meson and baryon spectra.
The multiquark Hamiltonian is taken to be
\begin{equation}
    H = \sum_{i=1}^{N} \left(m_i + \frac{p_i^2}{2m_i}\right)
      + \sum_{i<j} V_{ij},
    \label{eq:hamiltonian}
\end{equation}
where $m_i$ is the constituent quark mass.
The two-body interaction of the AL1 potential comprises the one-gluon-exchange interaction and linear confinement:
\begin{equation}
    V_{ij} = -\frac{3}{16}\,\boldsymbol\lambda_i\cdot\boldsymbol\lambda_j\,
    \Bigg[
        -\frac{\kappa}{r_{ij}} 
        + \frac{8\pi \kappa'}{3 m_i m_j}
          \frac{e^{-r_{ij}^2/r_0^2}}{\pi^{3/2} r_0^3}\!
          s_i\cdot s_j + \lambda\,r_{ij}- \Lambda
    \Bigg].
    \label{eq:al1_potential}
\end{equation}
Here $\boldsymbol\lambda_i$ denotes the Gell-Mann color matrices (with $i$ labeling the particle),
$\kappa$ and $\kappa'$ are the color-electric and color-magnetic coupling constants,
$\lambda$ is the string tension parameter,
$\Lambda$ is a zero-point energy constant,
and $r_0 = A\,[2m_i m_j/(m_i+m_j)]^{-B}$ is the characteristic smearing radius of the color-magnetic term.
Long-range confinement is modeled by the linear term $\lambda\,r_{ij}$.
The parameters of the AL1 potential model and the constituent quark masses are listed in Table~\ref{tab:al1_params}.

\begin{table*}[htbp]
    \centering
    \caption{Parameters of the AL1 potential model and the constituent quark masses (in GeV). The light quarks $u/d$ are mass-degenerate.}
    \label{tab:al1_params}
    \begin{tabular*}{\hsize}{@{}@{\extracolsep{\fill}}ccccccccccc@{}}
        \hline\hline
        $ \kappa $ &$ \lambda { [\mathrm{GeV}^{2}]}$&$ \Lambda {\rm [GeV]} $&$ \kappa^\prime $&$ A { [\mathrm{GeV}^{B-1}]}$&$ B $&$ m_b {\rm [GeV]}$&$ m_c {\rm [GeV]}$&$ m_s {\rm [GeV]}$&$ m_{u/d} {\rm [GeV]}$ \\
        \hline
        0.5069&0.1653&0.8321&1.8609&1.6553&0.2204&5.227&1.836&0.577&0.315\\
        \hline\hline
    \end{tabular*}
\end{table*}

\subsection{Construction of Discrete Degrees of Freedom}
\label{sec:discrete}

The total wave function of a multiquark system is defined on the product space of spatial coordinates and discrete degrees of freedom
(spin $\otimes$ isospin $\otimes$ color).

The spin subspace is labeled by the total spin $S$ and intermediate coupling quantum numbers,
constructed according to the SU(2) angular momentum coupling scheme.
The algebraic structure of the isospin subspace is entirely analogous,
except that only light quarks ($u$, $d$, and their antiparticles) belong to the two-dimensional representation of SU(2),
while $s$, $c$, and $b$ quarks are isospin singlets.

In the color subspace, 
color singlet basis states are constructed as follows. First, we enumerate all possible
baryon-type contractions ($\varepsilon^{abc} q_a q_b q_c$) 
and meson-type contractions ($\delta^a_b \, q_a \bar{q}^b$). These color singlet states are then extracted to obtain the maximally linearly independent set, and orthogonalized using the Gram--Schmidt procedure. For pentaquark states $q_1q_2q_3q_4\bar{q}$, there are 3 linearly independent color-singlet
bases, which can be chosen as
\begin{equation}
		\begin{aligned}
				\chi_{\bar 3_c\otimes\bar 3_c}=\left\{\left[(q_1 q_2)_{\bar 3_c}(q_3q_4)_{\bar 3_c}\right]_{3_c}\bar{q}\right\}_{1_c},\\
				\chi_{\bar 3_c\otimes 6_c}=\left\{\left[(q_1 q_2)_{\bar 3_c}(q_3 q_4)_{6_c}\right]_{3_c}\bar{q}\right\}_{1_c},\\
				\chi_{6_c\otimes\bar 3_c}=\left\{\left[(q_1 q_2)_{6_c}(q_3 q_4)_{\bar 3_c}\right]_{3_c}\bar{q}\right\}_{1_c}.\\
		\end{aligned}
		\label{eq:color_bases}
	\end{equation}

The direct product of the three subspaces above forms the total $D$-dimensional discrete space, with $D = d_{\text{spin}} \times d_{\text{iso}} \times d_{\text{color}}$, and $d_{\text{spin}}, d_{\text{iso}}, d_{\text{color}}$ are the dimensions of spin, isospin and color subspaces, respectively.

\subsection{Neural-network wave function}
The multiquark wave function is parametrized using the Neural-network Quantum States (NQS)~\cite{carleo2017}. The network architecture follows the general framework of DeepQuark~\cite{deepquark}, with several modifications introduced to improve computational efficiency and numerical accuracy.

The inputs of the network $x$ include the $3(N-1)$ Jacobi coordinates with the center-of-mass motion removed,
together with all inter-particle distances $r_{ij}$,
giving a total input dimension of $3(N-1) + N(N-1)/2$.
The network body consists of $N_{\text{layer}}=4$ fully connected layers, each containing $N_{\text{hid}}=32$ hidden neurons with the $\tanh$ activation function. In each layer, the hidden vectors are updated through
residual connections
$h^{(l+1)} = \tanh(W^{(l)} h^{(l)} + b^{(l)}) + h^{(l)}$, which alleviate vanishing gradients during the training~\cite{he2016resnet}. $\theta=\{W^{(l)} ,b^{(l)}\}$ are learnable parameters of the neural network. 
The output is a $D$-dimensional complex vector,
whose component $d^{\prime}$ gives the spatial wave function amplitude $f_{d^{\prime}}(x;\theta)$
of the discrete channel $d^{\prime}$ constructed in Sec.~\ref{sec:discrete}. Unlike our previous work~\cite{deepquark}, where the coupled-channel basis states were included as inputs to the network, here we employ a multi-head output architecture. This allows the amplitudes of all $D$ channels, $f_{d^{\prime}}(x;\theta)$, to be obtained in a single forward pass, instead of requiring $D$ separate evaluations, thereby improving computational efficiency.

To better describe the long-range behavior of the wave function, we multiply the network output by a learnable exponentially decaying envelope factor,
\begin{equation}
\psi_{d^{\prime}}(x;\theta)= f_{d^{\prime}}(x;\theta)\exp[-\sum_{i<j} (r_{ij}/b_{ij})^{3/2}].
\label{eq:envelope}
\end{equation}
The exponent $3/2$ is motivated by the asymptotic behavior of the wave function in a linear potential $V(r)\propto r$. The parameters $b_{ij}$ are initialized to $5~\mathrm{GeV}^{-1}$ and optimized together with the network parameters during training. Treating $b_{ij}$ as learnable parameters allows the envelope to adapt to the characteristic length scales of the system.

For systems containing identical fermions,
the antisymmetrized wave function is obtained by acting the antisymmetrization operator
on the envelope-modulated amplitude $\psi$,

\begin{equation}
\begin{aligned}
    \Psi_\theta(x,d)=&\left[\mathcal{A}\psi\right]_d(x;\theta)\\
    =&\sum_{\alpha=1}^{N_{\text{perm}}} \sum_{d^{\prime}=1}^{D}
      (U_\alpha)_{d d^{\prime}} \;
      \psi_{d^{\prime}}\!\bigl(P_\alpha^{-1} x; \theta\bigr).
    \label{eq:antisymmetrized_wf}
\end{aligned}
\end{equation}
Here $\mathcal{A} = \prod_{G} \sum_{P \in S(G)} \operatorname{sgn}(P) P$ is the antisymmetrization operator, where $G$ labels each group of identical quarks, and $S(G)$ is the permutation group of $G$. $(U_\alpha)_{d d^{\prime}}$ is the representation matrix of the permutation operator in the discrete space, and $P_\alpha^{-1} x$ denotes the rearrangement of spatial coordinates by the permutation operator. $\alpha$ sums over all elements of $\bigotimes_G S(G)$.
Moreover, we select states of definite parity via the parity projection
$\Psi_\theta^{P}(x,d) = (1+\pi \hat P)\Psi_\theta(x,d)$,
with $ \pi= \pm 1$ and $\hat P$ being the  spatial inversion operator.

\subsection{Variational Monte Carlo}
The neural network is trained in an unsupervised way based on the variational principle. By minimizing the energy expectation value
\begin{equation}
    E_\theta=\frac{\langle\Psi_\theta|H|\Psi_\theta\rangle}{\langle\Psi_\theta|\Psi_\theta\rangle}
\end{equation}
with respect to the network parameters $\theta$, we can get the ground-state solution represented by the NQS.
To estimate the energy expectation value and its gradient, we use the Markov Chain Monte Carlo sampling with the Metropolis--Hastings algorithm~\cite{metropolis1953,hastings1970} to obtain samples distributed by $|\Psi|^2$.
The sampled configuration includes both the spatial and discrete degrees of freedom $(x, d)$.
Each Metropolis move proposes a new state $(x^\prime,d^\prime)$. The spatial coordinates are generated by a Gaussian proposal $x^\prime= x+\delta$, where the displacement of particle $i$ is drawn from a Gaussian distribution $\delta_i\sim\mathcal{N}(0,\sigma_{f_i}^{2})$. The width $\sigma_{f}=\sigma\,r_{f}$ is flavor dependent, with $\sigma=0.10\;\mathrm{GeV}^{-1}$, $r_{u/d}=r_s=1.0$, $r_c=0.3$, $r_b=0.2$. This flavor-dependent scaling reduces the step size for heavier quarks, which keeps acceptance rates comparable across different flavor sectors.
 The discrete channel index $d^\prime$ is
drawn uniformly from the $D$ channels. The proposed state is accepted with probability $\beta=\min\!\left\{1,\;\dfrac{|\Psi_\theta(x^\prime,d^\prime)|^{2}}{|\Psi_\theta(x,d)|^{2}}\right\}$.

With the generated samples $\{(x,d)\}$, the energy expectation value $E_\theta$ and its gradient can be evaluated as
\begin{equation}
    \begin{aligned}
    &E_\theta =\mathbb{E}[E_L]\\
    &\frac{\partial E_\theta}{\partial \theta_k }= 2 \operatorname{Re} \mathbb{E}[(E_L - E_\theta) O_k^*]
    \end{aligned}
\end{equation}
where $\mathbb{E}[\cdot]$ is the Monte-Carlo average over $(x,d)$ drawn from $\propto|\Psi_\theta(x,d)|^{2}$,
and $O_k=\partial\ln\Psi_\theta/\partial\theta_k$.
The local energy is defined as
\begin{equation}
\begin{aligned}
&E_L(x, d) =
    \frac{\langle x, d \rvert H \lvert \Psi \rangle}
         {\Psi_\theta(x,d)} \\[0.7ex]
=& \frac{1}{\Psi_\theta(x,d)} \left[ -\sum_i \frac{\nabla_i^2}{2 m_i} \Psi_\theta(x,d) + \sum_{d^{\prime}=1}^{D} V_{d d^{\prime}}(x)\, \Psi_\theta(x,d^\prime) \right],
\end{aligned}
\label{eq:local_energy_full}
\end{equation}
where $V_{d d^{\prime}}(x)$ is the matrix element of the two-body potential
$\sum_{i<j} V_{ij}$.

Parameter optimization is carried out using the Stochastic Reconfiguration (SR) method~\cite{sorella2005},
which can be interpreted as natural-gradient descent~\cite{amari1998,sorella2001} on the variational wave-function manifold. The parameter update rule is
\begin{equation}
    \delta\theta = -\eta \, (\mathcal{M} + \lambda_{\text{reg}} 1)^{-1} \,
                   \nabla_\theta E,
    \label{eq:sr_update}
\end{equation}
where the real part of the quantum geometric tensor,
\begin{equation*}
    \mathcal{M}_{kl} = \operatorname{Re}\!\left\{
    \mathbb{E}[O_k^* O_l] - \mathbb{E}[O_k^*]\mathbb{E}[O_l]
    \right\},
\end{equation*}
is the Fubini--Study metric used for the real-parameter SR update, and $\lambda_{\text{reg}}$ is an adaptive regularization parameter. More details can be found in Appendix~\ref{sec:appendix_sr}. 
The numerical implementation is based on the JAX and NetKet frameworks~\cite{netket},
exploiting vectorized parallel computation of the wave function over all permutations and discrete channels,
with GPU acceleration through XLA just-in-time compilation.

\section{Unbound systems and scattering states}
\label{sec:results}


We first compute the ground state masses of all S-wave mesons and baryons, along with some P-wave baryons,
which are needed to determine the lowest baryon--meson thresholds of the pentaquark systems.
The hadron spectrum is summarized in Appendix Table~\ref{tab:hadrons_full}.
The results agree well with those calculated by the Gaussian expansion method~\cite{chen2026_excitedbaryon}, and the statistical errors in all calculations are below $0.1$~MeV. In general, the calculated energies are in good agreement with the experimental values. For light baryons, however, deviations of around 50 MeV are observed, which are due to our omission of the phenomenological three-body potential $V_{123}=-C/(m_1m_2m_3)$ in the AL1 model. Since this term contributes only a constant energy shift, it does not affect the pentaquark binding energies, which are determined by the energy differences between the pentaquark states and the corresponding baryon--meson thresholds.

For most pentaquark systems, no bound states below the lowest threshold are found. In the absence of bound states, the ground state is the baryon--meson scattering state at the lowest threshold. In our calculations, the converged energies of these unbound systems lie above the lowest baryon--meson thresholds, providing a baseline for identifying the existence of bound states. 

\begin{figure*}[htbp]
\centering
\includegraphics[width=\textwidth]{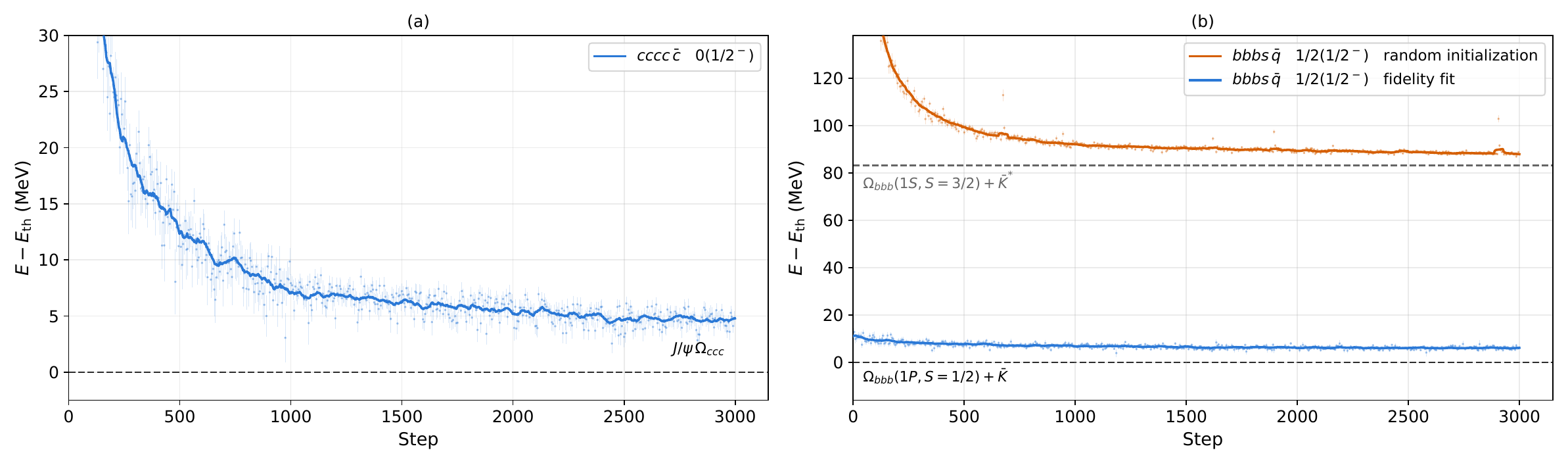}
\caption{Convergence of scattering states. The energy with respect to the lowest threshold
    $E-E_{\rm th}$ in MeV is plotted. The light scatter point with error bar
    is the energy at each step, while the solid lines are 41-point moving averages. 
    (a) For the $S=1/2$ fully charmed $cccc\cbar$ configuration, the lowest threshold
    is the S-wave $J/\psi\,\Omega_{ccc}$ channel, and the energy approaches
    it from above, converging to a plateau about 5~MeV above the threshold.
    Panel (b) shows the $bbbs\qbar$ system with $S=1/2$, where the
    lowest threshold is the P-wave $\Omega_{bbb}(1P,S=1/2)+\bar{K}$,
    indicated by the lower black dashed line, while the higher S-wave threshold
    $\Omega_{bbb}(1S,S=3/2)+\bar{K}^*$ is marked by the upper gray dashed line.
    The two curves show the optimization of the same system starting from two different initial wave functions.
    Starting from random initialization (orange), the optimization fails to reach the lowest threshold and
    converges instead to the S-wave threshold, ending on a plateau about
    87~MeV above the P-wave threshold and only a few MeV above the S-wave
    threshold.
    With the fidelity fit of Eq.~(\ref{eq:fidelity}) as the initial state
    (blue), the same system reaches the P-wave threshold instead,
    converging to a plateau about 6~MeV above it.}
\label{fig:pwave_scatter}
\end{figure*}

The ground states of the unbound systems can be classified into S-wave scattering states and P-wave scattering states. An S-wave scattering state consists of an S-wave meson and an S-wave baryon with relative angular momentum $L=0$. The NQS converges well to the S-wave scattering states at thresholds, as shown in Fig.~\ref{fig:pwave_scatter}(a) with the fully charmed pentaquark system as an example. The convergence pattern of other S-wave scattering states can be found in Figs.~\ref{fig:scattering_states} and \ref{fig:scattering_states_b} in Appendix~\ref{sec:appendix_scattering}.  The computed energies are a few MeV above the thresholds, which originate from the residual relative kinetic energy associated with the localized wave-function ansatz. The NQS converges to the $KN$ scattering state in the $qqqq\sbar $ channel as shown in Table~\ref{tab:scattering_states} and  Fig.~\ref{fig:scattering_states}(a), where there had been heated discussions on the existence of the $\Theta$ pentaquark. In $qqqc\cbar$ and $qqsc\cbar$ channels where the LHCb collaborations reported the hidden-charm pentaquark resonances, the NQS converges to the lowest $\eta_c N$ and $\eta_c \Lambda$ scattering states respectively as shown in Table~\ref{tab:scattering_states} and Fig.~\ref{fig:scattering_states}(c). It should be stressed that our ground-state calculations are not able to obtain the experimental resonant results. In fully heavy pentaquark systems, the NQS converges to the lowest heavy quarkonium and $\Omega_{QQQ}$ scattering states. We find that fully heavy pentaquarks with different flavor content and quantum numbers are all unbound, consistent with previous studies~\cite{an2022_fullyheavy,Gordillo:2024blx,li2026_fullyheavy}. 

On the other hand, the lowest threshold of some pentaquark systems consists of an S-wave meson and a P-wave baryon with relative angular momentum $L=1$, which we refer to as a P-wave scattering state. These systems include: 1) $S=\frac{1}{2}$ $bbb s \qbar$, $bbb q\sbar $ containing a P-wave spin $\frac{1}{2}$ $\Omega_{bbb}$ baryon; 2) $S=\frac{1}{2}$ $ccc q\sbar $ containing a P-wave spin $\frac{1}{2}$ $\Omega_{ccc}$ baryon; 3) $S=\frac{1}{2}$ $sss s\qbar $ containing a P-wave spin $\frac{1}{2}$ $\Omega$ baryon; 4) $I(S)=0(\frac{5}{2})$  $bsqq\bbar$ containing a P-wave spin $\frac{3}{2}$ $\Lambda$ baryon; 5) $I(S)=2(\frac{1}{2}$) $qqqq\cbar $ and $qqqq\sbar$ containing a P-wave $I(S)=\frac{3}{2}(\frac{1}{2})$ light baryon; 6) $I(S)=\frac{1}{2}(\frac{5}{2})$ $bqqq\bbar$, and $I(S)=0(\frac{5}{2})$ $qqqq \bbar$, $qqqq \cbar$, $qqqq \sbar$ containing a P-wave $I(S)=\frac{1}{2}(\frac{3}{2})$ light baryon.

For example, in the $bbbs\qbar$ system with $S=\frac{1}{2}$ and negative parity, the P-wave threshold $\Omega_{bbb}(1P,S=1/2)+\bar{K}$ lies below the lowest S-wave threshold $\Omega_{bbb}(1S,S=3/2)+\bar{K^*}$. However, we find that the NQS converges to the higher S-wave $\Omega_{bbb}(1S)\bar{K^*}$ scattering state rather than the lowest P-wave $\Omega_{bbb}(1P)\bar{K}$, as shown by the orange line in Fig.~\ref{fig:pwave_scatter}(b). This behavior is observed in several systems: starting from random initialization, the optimization tends to favor the simpler S-wave solution and can become trapped in the corresponding local minimum, instead of developing nontrivial spatial structure required for the P-wave state. This suggests a possible limitation of the present NQS optimization in resolving P-wave scattering states.  Nonetheless, the correct P-wave state can be recovered by providing the NQS with a suitable initialization. Specifically, we first fit the NQS to a P-wave packet,
\begin{equation}
	\Psi_p=(1+ \hat P)\mathcal A\left[
	U\Psi_B(\boldsymbol\xi_B)\Psi_M(\boldsymbol\xi_M)g_p(\boldsymbol{R})
	\right],
\end{equation}
where $\Psi_B$ and $\Psi_M$ are the $\Omega_{bbb}(1P)$ baryon and $\bar{K}$ meson wave functions, $\boldsymbol{R}$ is the relative distance between the baryon and meson center-of-mass coordinates, $g_p(R)=R\exp\left[-\frac{1}{2}(\frac{R}{\omega})^{3/2}\right]Y_{10}(\boldsymbol{R})$ is a P-wave packet, and $U$ is the color and spin coupling matrix. The fitting procedure is carried out by maximizing the fidelity,
\begin{equation}
    F=
  \frac{
    \left|\langle\Psi_p|\Psi_\theta\rangle\right|^2
  }{\langle\Psi_p|\Psi_p\rangle\langle\Psi_\theta|\Psi_\theta\rangle
  }\in[0,1].
\label{eq:fidelity}
\end{equation}
Then we use the fitted state as the initial wave function for the subsequent VMC optimization. With this initialization, the NQS successfully converges to the P-wave threshold, as shown by the blue line in Fig.~\ref{fig:pwave_scatter}(b). This further indicates that the NQS has the expressive capacity to represent the P-wave scattering state, while the main difficulty lies in reaching it from random initialization.

\section{Pentaquark Bound States}
\label{sec:bound_states}

\begin{figure*}[htbp]
\centering
\includegraphics[width=\textwidth]{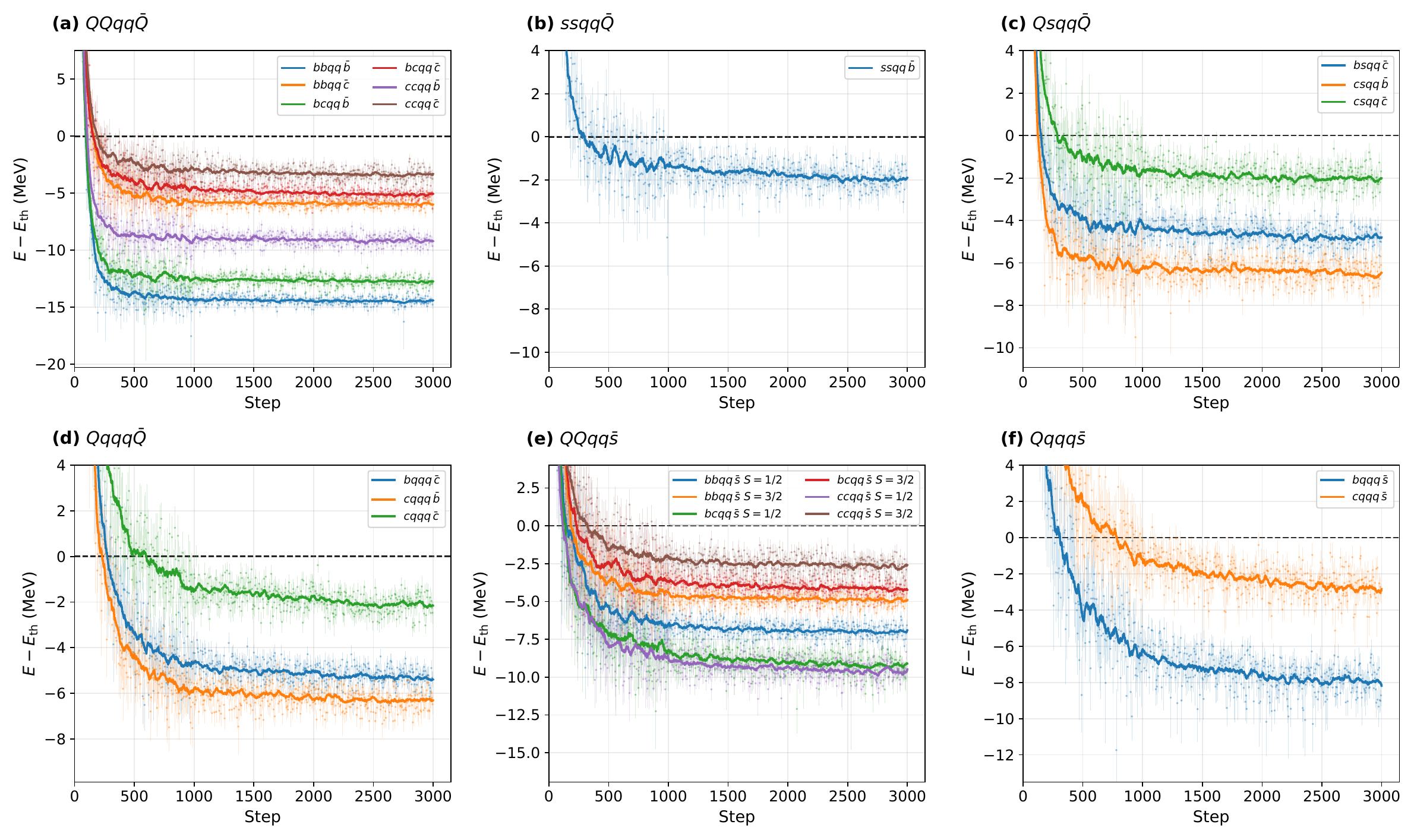}
\caption{
    Variational energies of the 21 bound states in the (a) $I(S^P)=0(\frac{5}{2}^-)$ $QQqq\bar{Q}$, (b) $I(S^P)=0(\frac{5}{2}^-)$ $ssqq\bar{Q}$, (c) $I(S^P)=0(\frac{5}{2}^-)$ $Qsqq\bar{Q}$, (d) $I(S^P)=\frac{1}{2}(\frac{5}{2}^-)$ $Qqqq\bar{Q}$, (e) $I(S^P)=0(\frac{1}{2}^-)$ and $0(\frac{3}{2}^-)$ $QQqq\bar{s}$, and (f) $I(S^P)=\frac{1}{2}(\frac{3}{2}^-)$ $Qqqq\bar{s}$ systems, as functions of iteration steps in the training process. The energy with respect to the lowest threshold
    $E-E_{\rm th}$ in MeV is plotted, and the dashed black lines indicate lowest baryon--meson thresholds. The light scatter point with error bar
    is the energy at each step, while the solid lines are 41-point moving averages. 
}
\label{fig:bound_states}
\end{figure*}

We identify 21 bound states in the $QQqq\bar{Q}$, $Qqqq\bar{Q}$, $Qsqq\bar{Q}$, $ssqq\bar{Q}$,  $QQqq\bar{s}$, and $Qqqq\bar{s}$ systems. Their binding energies with respect to the lowest threshold $E_b = E_{\text{th}} - E_{\text{ground}}$, proportions of color configurations and inter-quark root-mean-square (rms) radii are summarized in Table~\ref{tab:bound_states}. The optimization progress is shown in Fig.~\ref{fig:bound_states}, where the energies relative to the corresponding lowest baryon--meson thresholds are displayed. Notably, all of the bound states can be identified as shallow bound molecular states,
with binding energies ranging from $2$ to $15$ MeV.

\begin{table*}[htbp]
\centering
\caption{
    Binding energies, proportions of color configurations $\chi_{\bar{3}\otimes\bar{3}}$,
    $\chi_{\bar{3}\otimes 6}$, $\chi_{6\otimes\bar{3}}$, and inter-quark rms radii  $\langle r_{ij}^2\rangle^{1/2}$ of the pentaquark bound states.
    For the binding energies, the numbers in parentheses are statistical errors. The lowest baryon--meson threshold in each system is shown in the third column. 
}
\label{tab:bound_states}
{
\renewcommand{\arraystretch}{1.12}
\begin{tabular*}{\hsize}{@{}@{\extracolsep{\fill}}lccccccc@{}}
\hline\hline
System & $I(S^P)$ & Threshold & $E_b$ (MeV)
      & $\chi_{\bar{3}\otimes\bar{3}}$ (\%) & $\chi_{\bar{3}\otimes 6}$ (\%) & $\chi_{6\otimes\bar{3}}$ (\%) & $\langle r_{ij}^2\rangle^{1/2}$ (fm) \\
\hline
\multicolumn{8}{l}{\textbf{Group 1}: $QQqq\bar{Q}$ } \\
$bbqq\bbar$ & $0(5/2^-)$ & $ B^*\,\Xi_{bb}^*$ & 14.6(3) & 15.4 & 84.6 & 0.0 & $qq$ 1.32, $qb$ 0.96, $bb$ 0.31, $q\bbar$ 0.95, $b\bbar$ 1.02 \\
$bbqq\cbar$ & $0(5/2^-)$ & $\bar D^*\,\Xi_{bb}^*$ & 6.2(3) & 16.9 & 83.1 & 0.0 & $qq$ 1.60, $qb$ 1.16, $bb$ 0.30, $q\cbar$ 1.18, $b\cbar$ 1.39 \\
$bcqq\bbar$ & $0(5/2^-)$ & $ B^*\,\Xi_{cb}^*$ & 12.9(3) & 15.8 & 84.1 & 0.0 & $qq$ 1.34, $qb$ 0.98, $qc$ 1.01, $bc$ 0.41, $q\bbar$ 0.97, $c\bbar$ 1.10, $b\bbar$ 1.06 \\
$bcqq\cbar$ & $0(5/2^-)$ & $\bar D^*\,\Xi_{cb}^*$ & 5.3(4) & 17.5 & 82.4 & 0.0 & $qq$ 1.64, $qb$ 1.18, $qc$ 1.20, $bc$ 0.42, $q\cbar$ 1.20, $c\cbar$ 1.43, $b\cbar$ 1.41 \\
$ccqq\bbar$ & $0(5/2^-)$ & $ B^*\,\Xi_{cc}^*$ & 9.3(4) & 15.8 & 84.1 & 0.0 & $qq$ 1.47, $qc$ 1.09, $cc$ 0.50, $q\bbar$ 1.05, $c\bbar$ 1.22 \\
$ccqq\cbar$ & $0(5/2^-)$ & $\bar D^*\,\Xi_{cc}^*$ & 3.7(4) & 18.7 & 81.2 & 0.1 & $qq$ 1.82, $qc$ 1.32, $cc$ 0.50, $q\cbar$ 1.32, $c\cbar$ 1.64 \\
\hline
\multicolumn{8}{l}{\textbf{Group 2}: $ssqq\bar{Q}$ } \\
$ssqq\bbar$ & $0(5/2^-)$ & $\bar B^*\,\Xi^*$ & 2.4(4) & 17.8 &  82.1 & 0.1 & $qq$ 2.01, $qs$ 1.53, $ss$ 0.79, $q\bbar$ 1.44, $s\bbar$ 1.88 \\
\hline
\multicolumn{8}{l}{\textbf{Group 3}: $Qsqq\bar{Q}$ } \\
$bsqq\cbar$ & $0(5/2^-)$ & $\bar D^*\,\Xi_b^*$ & 5.0(4) & 17.4 & 82.5 & 0.1 & $qq$ 1.61, $qs$ 1.26, $qb$ 1.16, $bs$ 0.61, $q\cbar$ 1.18, $s\cbar$ 1.49, $b\cbar$ 1.38 \\
$csqq\bbar$ & $0(5/2^-)$ & $ B^*\,\Xi_c^*$ & 6.7(5) & 16.3 & 83.6 & 0.1 & $qq$ 1.53, $qs$ 1.21, $qc$ 1.14, $cs$ 0.66, $q\bbar$ 1.10, $s\bbar$ 1.36, $c\bbar$ 1.29 \\
$csqq\cbar$ & $0(5/2^-)$ & $\bar D^*\,\Xi_c^*$ & 2.2(4) & 19.1 & 80.8 & 0.1 & $qq$ 1.91, $qs$ 1.45, $qc$ 1.40, $cs$ 0.67, $q\cbar$ 1.39, $s\cbar$ 1.78, $c\cbar$ 1.73 \\
\hline
\multicolumn{8}{l}{\textbf{Group 4}: $Qqqq\bar{Q}$ } \\
$bqqq\cbar$ & $1/2(5/2^-)$ & $\bar D^*\,\Sigma_b^*$ & 5.5(4) & 33.1 & 49.2 & 17.7 & $qq$ 1.41, $qb$ 1.04, $q\cbar$ 1.30, $b\cbar$ 1.36 \\
$cqqq\bbar$ & $1/2(5/2^-)$ & $ B^*\,\Sigma_c^*$ & 6.6(4) & 33.2 & 48.1 & 18.7 & $qq$ 1.41, $qc$ 1.06, $q\bbar$ 1.26, $c\bbar$ 1.34 \\
$cqqq\cbar$ & $1/2(5/2^-)$ & $\bar D^*\,\Sigma_c^*$ & 2.4(4) & 33.4 & 45.7 & 20.9 & $qq$ 1.53, $qc$ 1.15, $q\cbar$ 1.42, $c\cbar$ 1.55 \\
\hline
\multicolumn{8}{l}{\textbf{Group 5}: $QQqq\bar{s}$} \\
$bbqq\sbar$ & $0(1/2^-)$ & $ K\,\Xi_{bb}$ & 7.5(4) & 32.2 & 67.7 & 0.1 & $qq$ 1.62, $qb$ 1.17, $bb$ 0.31, $q\sbar$ 1.20, $b\sbar$ 1.49 \\
$bbqq\sbar$ & $0(3/2^-)$ & $ K\,\Xi_{bb}^*$ & 5.3(4) & 31.5 & 68.4 & 0.1 & $qq$ 1.64, $qb$ 1.19, $bb$ 0.32, $q\sbar$ 1.21, $b\sbar$ 1.51 \\
$bcqq\sbar$ & $0(1/2^-)$ & $ K\,\Xi_{cb}$ & 10.0(6) & 32.7 & 66.2 & 1.1 & $qq$ 1.40, $qb$ 1.02, $qc$ 1.05, $bc$ 0.41, $q\sbar$ 1.05, $c\sbar$ 1.28, $b\sbar$ 1.25 \\
$bcqq\sbar$ & $0(3/2^-)$ & $ K\,\Xi_{cb}^*$ & 4.6(5) & 31.4 & 68.5 & 0.1 & $qq$ 1.76, $qb$ 1.28, $qc$ 1.31, $bc$ 0.43, $q\sbar$ 1.30, $c\sbar$ 1.68, $b\sbar$ 1.64 \\
$ccqq\sbar$ & $0(1/2^-)$ & $ K\,\Xi_{cc}$ & 10.3(5) & 32.7 & 66.2 & 1.1 & $qq$ 1.47, $qc$ 1.10, $cc$ 0.49, $q\sbar$ 1.11, $c\sbar$ 1.36 \\
$ccqq\sbar$ & $0(3/2^-)$ & $ K\,\Xi_{cc}^*$ & 3.1(5) & 31.5 & 68.4 & 0.1 & $qq$ 2.02, $qc$ 1.46, $cc$ 0.50, $q\sbar$ 1.47, $c\sbar$ 1.89 \\
\hline
\multicolumn{8}{l}{\textbf{Group 6}: $Qqqq\bar{s}$} \\
$bqqq\sbar$ & $1/2(3/2^-)$ & $ K\,\Sigma_b^*$ & 8.5(6) & 33.3 & 40.1 & 26.6 & $qq$ 1.34, $qb$ 0.99, $q\sbar$ 1.25, $b\sbar$ 1.31 \\
$cqqq\sbar$ & $1/2(3/2^-)$ & $ K\,\Sigma_c^*$ & 3.4(7) & 33.4 & 41.9 & 24.7 & $qq$ 1.44, $qc$ 1.09, $q\sbar$ 1.35, $c\sbar$ 1.48 \\

\hline\hline
\end{tabular*}}
\end{table*}

\subsection{$QQqq\bar{Q}$, $ssqq\bar{Q}$, $Qsqq\bar{Q}$ and $Qqqq\bar{Q}$}

In the $QQqq\bar{Q}$, $ssqq\bar{Q}$, $Qsqq\bar{Q}$, and $Qqqq\bar{Q}$ systems, we identify six, one, three bound states with $I(S^P)=0(\frac{5}{2}^-)$ and three bound states with
$I(S^P)=\frac{1}{2}(\frac{5}{2}^-)$, respectively. These bound states share a very similar underlying dynamics, which can be seen from their color wave functions and internal structures. In the $QQqq\bar{Q}$, $Qsqq\bar{Q}$, and $ssqq\bar{Q}$ systems, the color configurations are dominated by the color-antitriplet diquark $(QQ)_{\bar 3_c}$, $(Qs)_{\bar 3_c}$ and $(ss)_{\bar 3_c}$. As the reduced mass of the diquark decreases, their size increases. From Table~\ref{tab:bound_states}, we observe a very clear pattern: $r_{bb} < r_{bc} < r_{cc} < r_{bs} < r_{cs} < r_{ss} $. 
In the $Qqqq\bar{Q}$ system, all three light quarks are identical and there does not exist a compact $(Qq)$ core, leading to an appreciable $\chi_{6\otimes\bar{3}}$ component of its color wave function.

We take the $QQqq\bar{Q}$ system as an example.  From the color compositions and rms radii listed in Table~\ref{tab:bound_states}, we find that the two heavy quarks form a compact color antitriplet diquark $(QQ)_{\bar 3_c}$ with a size of around $0.3\text{--}0.5$ fm, characteristic of the compact $QQ$ core in a doubly heavy baryon. On the other hand, the distance $r_{Q\bar{Q}}$ and $r_{qq}$ are considerably larger, being above $1$ fm. This is consistent with a predominant baryon--meson molecular configuration composed of a heavy meson ${B}^*/\bar{D}^*$ and a doubly heavy baryon $\Xi_{QQ}^*$. Similarly, the $ssqq\bbar$ bound state is a ${B}^*\Xi^*$ molecule, the $Qsqq\bar{Q}$ bound states are ${B}^*/\bar{D}^*+\Xi_b^*/\Xi_c^*$ molecules, and the $Qqqq\bar{Q}$ bound states are ${B}^*/\bar{D}^*+\Sigma_b^*/\Sigma_c^*$ molecules. The $bsqq\bbar$ and $bqqq\bbar$ bound states are not identified, because the $\Upsilon$ meson is strongly bound and the lowest threshold in these systems are the P-wave $\Upsilon \Lambda^*$ and $\Upsilon N^*$, respectively.

Notably, the $S=\frac{5}{2}$ $qqqc\cbar$ and $qqsc\cbar$ bound states are identified as $\bar{D}^*\Sigma_c^*$ and $\bar{D}^*\Xi_c^*$ molecules, respectively. 
They have also been discussed in previous studies~\cite{Wang:2015qlf,Yang:2015bmv,Wu:2017weo,Meng:2019ilv,Wang:2019ato,Wang:2019nvm,Chen:2019asm,Du:2019pij,Liu:2019tjn,Chen:2019bip,Wang:2019spc,Weng:2019ynv,Giachino:2022pws,Lin:2023ihj,Wang:2023eng,wang2024_tripleheavy,Liang:2026penta} and can be considered as the partners of the experimental $P_{c\cbar}$ and $P_{c\cbar s}$ states within the heavy quark spin symmetry. 
Although these experimental resonances cannot be directly accessed in our ground-state calculations, the existence of their heavy-quark-spin partners provides an indirect glimpse into the underlying molecular dynamics and supports a common molecular interpretation of these states.

Considering the heavy diquark-antiquark symmetry, the compact diquark $ (QQ)_{\bar 3_c}$ is analogous to a heavy antiquark.
One may wonder whether the $QQqq\bar{Q}$ pentaquark bound states can be viewed as the analogues of the doubly heavy tetraquark states $\bar{Q}\bar{Q}qq$,  such as the deeply bound $T_{bb}$ and the molecular $T_{cc}$. However, in our calculations, only the molecular bound states are obtained while the deeply bound analogues remain absent. There is a clear difference between their light diquarks, which leads to the non-existence of deeply bound pentaquark states.

The color configuration of  $T_{bb}$ is dominated by $ \left[(\bbar\bbar)_{3_c}(qq)_{\bar3_c}\right]_{1_c}$, where the two bottomed antiquarks form a color-triplet source~\cite{Wu:2024doublyheavy}. The isospin and spin wave functions of the light quark pair are anti-symmetric while their orbital wave function is symmetric. They are the so-called good diquark~\cite{Jaffe:2004ph}, which is essential for the formation of deeply bound state. However,  the good diquark configuration is not allowed in the 
the $I(S^P)=0(\frac{5}{2}^-)$ $QQqq\bar{Q}$ pentaquark systems. For S-wave configurations, the light quark pair can only couple to $(qq)_{6_c}^{S=1,I=0}$, which belong to the so-called bad diquark with a higher mass and less binding than the good diquark. As shown in Table \ref{tab:bound_states}, ${6_c}$ is the dominant color wave function of the light quark pair. It is also interesting to note that there is still around $16\%$ $(qq)_{\bar 3_c}$ color configuration, which requires an anti-symmetric orbital wave function. Such a light diquark with an internal P-wave excitation clearly has an even higher mass than both the good and bad diquarks. Throughout our calculation, we ensure the total parity of the pentaquark state to be negative. Hence there also exists a P-wave orbital excitation between the four-quark cluster and heavy anti-quark. In short summary, both color configurations of the light diquark render the above six $QQqq\bar{Q}$ pentaquark states with $I(S^P)=0(\frac{5}{2}^-)$ to be weakly bound. The potential deeply bound $QQqq\bar{Q}$ pentaquark states as the analogue of $T_{bb}$ shall contain a good diquark, which can be realized in $S^P=\frac{1}{2}^-$ or $\frac{3}{2}^-$ systems. Unfortunately, in these systems the good diquarks are formed within the $\Lambda_b$ baryon, and the lowest states are $\eta_b \Lambda_b$ or 
$\Upsilon \Lambda_b$ scattering states.

For these weakly bound states, we observe a clear mass dependence of the binding energies. In the molecular picture, the relative kinetic energy of the baryon--meson system is governed by the reduced mass $ \mu=M_M M_B/(M_M+M_B)$, where $M_M$ and $M_B$ denote the masses of the heavy meson and doubly heavy baryon, respectively. For a comparable inter-hadron interaction, a larger reduced mass suppresses the relative kinetic energy and therefore favors stronger binding. For a fixed heavy meson such as ${B}^*$, the increasing baryon mass leads to a larger reduced mass, consistent with the binding-energy ordering $ E_b(ssq q\bbar)< E_b(cqq q\bbar)<E_b(csq q\bar b)<E_b(ccqq\bbar)<E_b(bcqq\bbar)<E_b(bbq q\bbar)$.
Likewise, for a fixed baryon, replacing $\bar D^*$ by the heavier $B^*$ increases the reduced mass, leading to a systematically larger binding energy. 

Although these pentaquark states lie below their corresponding $S$-wave $J=\frac{5}{2}$ baryon--meson thresholds, they may still decay into lower-lying two-body baryon--meson channels through higher partial waves. For example, the $cqqq\cbar$ bound state may decay via D-wave into the $\eta_c N$ and $J/\psi N$ channels. Such two-body decays may have a strongly suppressed partial width due to the $D$-wave centrifugal barrier. 
In contrast, their three-body (or four-body) decay widths may be larger, since the constituent  vector mesons and baryons decay easily. $\bar D^*$ decays into $\bar D \pi$ and $\bar D \gamma$. $B^*$ mainly decays into $B\gamma$. $\Xi^*$ decays into $\Xi \pi$ and $\Xi \gamma$. $\Sigma_Q^*$ decays into $\Lambda_Q \pi$. $\Xi_Q^*$ and $\Xi_{QQ}^*$ mainly decays to $\Xi_Q\pi$ and $\Xi_{QQ}\gamma$, respectively.

\subsection{$QQqq\bar{s}$ and $Qqqq\bar{s}$}

In the $QQqq\sbar$ system, we identify three bound states with $I(S^P)=0(\frac{1}{2}^-)$ and $0(\frac{3}{2}^-)$ respectively. The two heavy quarks form a compact diquark $(QQ)_{\bar 3_c}$ with a size of around $0.3\text{--}0.5$ fm, characteristic of the compact $QQ$ core in a doubly heavy baryon. The distance $r_{Q\bar{s}}$ lies within [1.25, 1.89] fm and $r_{qq}$ is in the range [1.40, 2.02] fm, which is consistent with a predominant molecular configuration composed of a $K$ meson and a doubly heavy baryon $\Xi_{QQ}^{(*)}$. Considering the heavy diquark-antiquark symmetry, the $QQqq\sbar$ bound states are the analogues of the $\bar{Q}\sbar qq$ bound states~\cite{chen2024_unified}. It is worth mentioning that lattice calculations also support attractive interactions in the isoscalar $\Xi_{cc}K$ channel, but a virtual state instead of a bound state was obtained in Ref.~\cite{Yi:2025lattice}.

The compact $(QQ)_{\bar 3_c}$ diquark core acts like an anti-heavy quark 
$\bar Q^{\prime}$ with a mass of $2m_Q$. If the $\bar s$ quark mass were as heavy as the $\bar Q^{\prime}$, $\bar s$ and $\bar Q^{\prime}$ might form a very compact anti-diquark $(\bar Q^{\prime} \bar s)_{3_c}$, which would behave like a very heavy quark denoted as $Q^{\prime\prime}$. Combined with the good diquark configuration $(qq)_{3_c}^{S=0,I=0}$ allowed in the $S=\frac{1}{2}$ and $\frac{3}{2}$ systems, $Q^{\prime\prime} qq \,(\Lambda_{Q^{\prime\prime}})$ would be the exact analogue of the deeply bound $T_{bb}$. However, in reality, the strange quark mass is much smaller than the charm and bottom quark mass, and the above assumption does not hold. The ratio between the $\bar3_c$ and $6_c$ light diquark is around $1:2$ from Table~\ref{tab:bound_states}.   Even if the strange quark were heavy enough, the threshold energy of the heavy quarkonium and $\Lambda_{Q}$ would be still lower than the $\Lambda_{Q''}$ mass, as we have seen in the $QQqq\bar{Q}$ system, and $\Lambda_{Q''}$ would not be a deeply bound state. 

In the $Qqqq\sbar$ system, we only find two bound states with  $I(S^P)=\frac{1}{2}(\frac{3}{2}^-)$. They are $K\Sigma_Q^*$ molecular states. Similar to the $Qqqq\bar{Q}$ system, a compact diquark is absent and there is an appreciable $\chi_{6\otimes\bar{3}}$ component. 

The fact that only the $S=\frac{3}{2}$  systems are bound while $S=\frac{1}{2}$ and $\frac{5}{2}$ systems are not, can be understood in terms of the good diquark picture. In the ${K}\Sigma_Q^*$ scattering state, the two light quarks in $\Sigma_Q^*$ form a $(qq)_{3_c}^{S=1,I=1}$ diquark, while the light quark in the $K$ meson is uncorrelated with those in $\Sigma_Q^*$. However, in the  $S=\frac{3}{2}\,Qqqq\sbar$ system, the light quarks can rearrange to form the good diquark configuration $(qq)_{3_c}^{S=0,I=0}$, which provide additional attraction and results in a pentaquark bound state. By contrast, in the $S=\frac{1}{2}$ system, the lowest threshold $K\Lambda_Q$ already consists of the good diquark and no extra binding can be formed to support a bound state below the threshold. In the $S=\frac{5}{2}$ system, the good diquark configuration is forbidden by the spin coupling.  


The spin $\frac{1}{2}$ $QQqq\sbar$ bound states are stable against strong interaction. They can be searched for in the weak decay process. On the other hand, the spin $\frac{3}{2}$ $cqqq\sbar$ and $bqqq\sbar$ bound state may be searched for in the $\Lambda_cK\pi$ and $\Lambda_bK\pi$ channels, respectively. Radiative decays are the dominant decay modes of all the other spin $\frac{3}{2}$ states. 

The $QQqq\bar{s}$ and $Qqqq\bar{s}$ bound states are composed of a pseudoscalar $K$ meson and heavy baryon. In contrast, the $QQqq\bar{Q}$, $ssqq\bar{Q}$, $Qsqq\bar{Q}$ and $Qqqq\bar{Q}$ bound states are loosely bound molecular states composed of a heavy vector meson and baryon. 
Within the framework of chiral perturbation theory, the heavy vector meson and baryon are matter fields. The pseudoscalar meson interacts with the matter field directly~\cite{Meng:2018zbl,Huang:2021fdt}, while two matter fields interact through the exchange of pseudoscalar mesons~\cite{Wang:2020dhf,Meng:2019nzy}. Our framework presents a unified and consistent description of the above two different pictures at the quark level successfully.

\section{Summary}
\label{sec:conclusion}

We have carried out a systematic computational study of the ground-state
properties of pentaquark systems using the NN-VMC method within the AL1 constituent quark potential model.
We scan over 387 manifestly exotic pentaquark configurations covering all possible color, spin, and flavor combinations, and identify 21 pentaquark bound state candidates
with binding energies ranging from $2$ to $15$~MeV. They are all baryon--meson molecular states. Specifically, the $QQqq\bar{Q}, ssqq\bar{Q}, Qsqq\bar{Q}, Qqqq\bar{Q}$ bound states with spin $\frac{5}{2}$ consist of a heavy vector meson and a baryon, while the $QQqq\sbar$ and $Qqqq\sbar$ bound states with spin $\frac{1}{2}$ or $\frac{3}{2}$ consist of a $K$ meson and a heavy baryon. Notably, a deeply bound compact configuration is absent in the pentaquark system, in contrast to the $T_{bb}$ state in the tetraquark sector.

The main decay modes of the spin $\frac{5}{2}$ $QQqq\bar{Q}$, $ssqq\bar{Q}$, $Qsqq\bar{Q}$, and $Qqqq\bar{Q}$ bound states may involve three-body final states. For example, the $cqqq\cbar$ bound state with $\bar{D}^*\Sigma_c^*$ molecular configuration can decay into the $\bar{D}^*\Lambda_c\pi$ channel via P-wave. Similarly, the spin $\frac{3}{2}$ $bqqq\sbar$ and $cqqq\sbar$ bound states can decay into the $\Lambda_bK\pi$ and $\Lambda_cK\pi$ channels, respectively. On the other hand, the spin $\frac{3}{2}$ $QQqq\sbar$ bound states decay radiatively, while the spin $\frac{1}{2}$ $QQqq\sbar$ bound states can only decay weakly. These pentaquark bound states may be searched for at LHCb, Belle~II, and other experiments in the future.

All of the bound states take the lowest isospin quantum number $I=0$ or $\frac{1}{2}$. Compared with higher isospin states, the lowest isospin configurations generally contain more good light diquark components, which are favored by the color-magnetic interactions.

As a cross-validation, the ground-state energies of unbound systems approaches the baryon--meson threshold closely from above,
consistent with the S-wave scattering states. In particular, we reproduce the lowest scattering states $\eta_c N$ and $\eta_c \Lambda$ in the $qqqc\cbar$ and $qqsc\cbar$ channels respectively, where the hidden-charm pentaquark resonances were observed experimentally. We also find that all fully heavy pentaquark systems are unbound. However, in some systems, the lowest threshold is the P-wave scattering state. Starting from a randomly initialized configuration, the NQS tends to converge to a simpler S-wave scattering state and fail to develop the correct P-wave structure. This can be remedied by providing the NQS a suitable initialization. 


The flexibility and scalability of the NN-VMC method make it well suited for extension to more complex systems such as the hexaquark systems.
The method can also be generalized to incorporate more sophisticated quark interactions, which may yield valuable insights into
the internal structure of multiquark states and nonperturbative Quantum
Chromodynamics. 


\section*{Acknowledgments}
\label{sec:acknowledgments}
This project was supported by the National Natural Science Foundation of China (Grant No. 12475137 and No.125B2105). The computational resources were supported by the High-Performance Computing Platform of Peking
University.

\appendix
\section{Hadron Ground State Energies}
\label{sec:appendix_hadrons}

Table~\ref{tab:hadrons_full} lists the ground-state energies of all mesons ($q\bar{q}$)
and baryons ($qqq$) computed with the AL1 potential model~\cite{silvestre1994,silvestre1996}
using the neural-network method of this work, together with the corresponding
experimental values from the PDG. The statistical error of each calculation
is below $0.1$~MeV.

\begin{table*}[htbp]
\centering
\caption{
    Meson and baryon ground-state energies (in MeV) in the AL1 model, computed with the NN-VMC
    method. Experimental values from the PDG~\cite{pdg2026} are listed for comparison where available. For the isospin multiplets,
    the experimental masses are taken to be the isospin average.
    The $\eta$ and $\eta'$ mesons are not listed as flavor mixing  is not considered in the current framework. 
}
\label{tab:hadrons_full}
\begin{tabular*}{\hsize}{@{}@{\extracolsep{\fill}}lccc@{\quad}lccc@{}}
\hline\hline
State & $I(S^P)$ & NN-VMC (MeV) & Expt. (MeV) & State & $I(S^P)$ & NN-VMC (MeV) & Expt. (MeV) \\
\hline
$\pi\,(q\bar{q})$     & $1(0^-)$   & 138.2 & 138.1 & $\rho\,(q\bar{q})$       & $1(1^-)$  & 770.0 & 775.3 \\
$\omega\,(q\bar{q})$ & $0(1^-)$  & 770.0 & 782.7 & $\phi\,(s\bar s)$    & $0(1^-)$   & 1020.8 & 1019.5\\
$\bar K\,(s\bar{q})$  & $1/2(0^-)$ & 491.0 & 495.6 & $\bar K^*\,(s\bar{q})$   & $1/2(1^-)$ & 903.5 & 893.6 \\
 $D\,(c\bar{q})$         & $1/2(0^-)$ & 1862.4 & 1867.3 &
$D^*\,(c\bar{q})$     & $1/2(1^-)$ & 2016.1 & 2008.6 \\
$D_s\,(c\bar s)$       & $0(0^-)$  & 1962.5 & 1968.3 & $D_s^*\,(c\bar s)$   & $0(1^-)$   & 2102.1 & 2112.2 \\ $\eta_c\,(c\bar c)$    & $0(0^-)$  & 3005.3 & 2983.9 &
$J/\psi\,(c\bar c)$  & $0(1^-)$   & 3101.3 & 3096.9 \\ $\bar B\,(b\bar{q})$     & $1/2(0^-)$ & 5293.5 & 5279.5 &
$\bar B^*\,(b\bar{q})$ & $1/2(1^-)$ & 5350.5 & 5325.0 \\ $\bar B_s\,(b\bar s)$   & $0(0^-)$  & 5361.0 & 5366.9 &
$\bar B_s^*\,(b\bar s)$ & $0(1^-)$ & 5417.5 & 5415.4 \\ $\bar B_c\,(b\bar c)$   & $0(0^-)$  & 6291.6 & 6274.5 &
$\bar B_c^*\,(b\bar c)$ & $0(1^-)$ & 6343.2 & --- \\ $\eta_b\,(b\bar b)$     & $0(0^-)$  & 9423.7 & 9398.7 &
$\Upsilon\,(b\bar b)$ & $0(1^-)$  & 9461.5 & 9460.3 \\
\hline
$N\,(qqq)$            & $1/2(1/2^+)$ & 995.1   & 938.9  & $N^*\,(qqq)$            & $1/2(3/2^-)$ & 1675.2  & --- \\
$\Delta\,(qqq)$       & $3/2(3/2^+)$ & 1306.6  & 1232.0 & $\Delta^*\,(qqq)$       & $3/2(1/2^-)$ & 1560.1  & --- \\
$\Lambda\,(qqs)$      & $0(1/2^+)$   & 1150.2  & 1115.7 & $\Lambda^*\,(qqs)$      & $0(3/2^-)$   & 1819.2  & --- \\
$\Sigma\,(qqs)$       & $1(1/2^+)$   & 1229.2  & 1193.1 & $\Sigma^*\,(qqs)$       & $1(3/2^+)$   & 1437.0  & 1383.7 \\
$\Xi\,(qss)$          & $1/2(1/2^+)$ & 1340.1  & 1314.9 & $\Xi^*\,(qss)$          & $1/2(3/2^+)$ & 1559.1  & 1533.4 \\
$\Omega\,(sss)$       & $0(3/2^+)$   & 1674.7  & 1672.5 & $\Omega^*\,(sss)$       & $0(1/2^-)$   & 1978.6  & --- \\
$\Lambda_c\,(qqc)$    & $0(1/2^+)$   & 2290.1  & 2286.5 & $\Lambda_c^*\,(qqc)$    & $0(3/2^-)$   & 2949.2  & --- \\
$\Sigma_c\,(qqc)$     & $1(1/2^+)$   & 2464.9  & 2453.5 & $\Sigma_c^*\,(qqc)$     & $1(3/2^+)$   & 2545.6  & 2518.1 \\
$\Xi_c\,(qsc)$        & $1/2(1/2^+)$ & 2469.5  & 2469.9 & $\Xi_c^*\,(qsc)$        & $1/2(3/2^+)$ & 2651.0  & 2645.5 \\
$\Omega_c\,(ssc)$     & $0(1/2^+)$   & 2677.2  & 2695.2 & $\Omega_c^*\,(ssc)$     & $0(3/2^+)$   & 2751.9  & 2765.9 \\
$\Xi_{cc}\,(qcc)$     & $1/2(1/2^+)$ & 3607.7  & 3621.6 & $\Xi_{cc}^*\,(qcc)$     & $1/2(3/2^+)$ & 3701.8  & --- \\
$\Omega_{cc}\,(scc)$  & $0(1/2^+)$   & 3708.6  & ---    & $\Omega_{cc}^*\,(scc)$  & $0(3/2^+)$   & 3791.3  & --- \\
$\Omega_{ccc}\,(ccc)$ & $0(3/2^+)$   & 4798.7  & ---    & $\Omega_{ccc}^*\,(ccc)$ & $0(1/2^-)$   & 5084.7  & --- \\
$\Lambda_b\,(qqb)$    & $0(1/2^+)$   & 5635.3  & 5619.6 & $\Lambda_b^*\,(qqb)$    & $0(3/2^-)$   & 6292.1  & --- \\
$\Sigma_b\,(qqb)$     & $1(1/2^+)$   & 5847.6  & 5811.6 & $\Sigma_b^*\,(qqb)$     & $1(3/2^+)$   & 5878.2  & 5832.4 \\
$\Xi_b\,(qsb)$        & $1/2(1/2^+)$ & 5803.8  & 5794.5 & $\Xi_b^*\,(qsb)$        & $1/2(3/2^+)$ & 5973.3  & 5955.3 \\
$\Omega_b\,(ssb)$     & $0(1/2^+)$   & 6033.6  & 6046.1 & $\Omega_b^*\,(ssb)$     & $0(3/2^+)$   & 6064.2  & --- \\
$\Xi_{cb}\,(qcb)$     & $1/2(1/2^+)$ & 6914.3  & ---    & $\Xi_{cb}^*\,(qcb)$     & $1/2(3/2^+)$ & 6990.1  & --- \\
$\Omega_{cb}\,(scb)$  & $0(1/2^+)$   & 7002.1  & ---    & $\Omega_{cb}^*\,(scb)$  & $0(3/2^+)$   & 7070.4  & --- \\
$\Omega_{ccb}\,(ccb)$ & $0(1/2^+)$   & 8017.3  & ---    & $\Omega_{ccb}^*\,(ccb)$ & $0(3/2^+)$   & 8046.1  & --- \\
$\Xi_{bb}\,(qbb)$     & $1/2(1/2^+)$ & 10193.6 & ---    & $\Xi_{bb}^*\,(qbb)$     & $1/2(3/2^+)$ & 10232.4 & --- \\
$\Omega_{bb}\,(sbb)$  & $0(1/2^+)$   & 10266.1 & ---    & $\Omega_{bb}^*\,(sbb)$  & $0(3/2^+)$   & 10303.5 & --- \\
$\Omega_{cbb}\,(cbb)$ & $0(1/2^+)$   & 11215.1 & ---    & $\Omega_{cbb}^*\,(cbb)$ & $0(3/2^+)$   & 11247.1 & --- \\
$\Omega_{bbb}\,(bbb)$ & $0(3/2^+)$   & 14398.2 & ---    & $\Omega_{bbb}^*\,(bbb)$ & $0(1/2^-)$   & 14727.7 & --- \\
\hline\hline
\end{tabular*}
\end{table*}

\section{Details of the Variational Monte Carlo Method}
\label{sec:appendix_sr}

\subsection{Variational Principle and Energy Functional}

Consider a trial wave function $\Psi_\theta(x)$ parameterized by $N_p$ real parameters $\theta = (\theta_1, \ldots, \theta_{N_p})^\top$.
The variational energy is defined as the Rayleigh quotient
\begin{equation}
    E_\theta = \frac{\langle \Psi_\theta | H | \Psi_\theta \rangle}
                     {\langle \Psi_\theta | \Psi_\theta \rangle}
            = \frac{\int dx\; \Psi_\theta^*(x)\, H \Psi_\theta(x)}
                     {\int dx\; |\Psi_\theta(x)|^2}.
    \label{eq:rayleigh}
\end{equation}
By the variational principle, $E_\theta \geq E_0$ for any trial wave function, with equality if and only if $\Psi_\theta$ coincides with the exact ground state wave function. The ground state problem is thus reduced to a continuous optimization problem in parameter space: $\min_{\theta} E_\theta$.
This variational principle lays the theoretical foundation of the neural network variational Monte Carlo (NN-VMC) method.

\subsection{Energy Gradient and the Force Formula}

The partial derivative of the energy with respect to the $k$-th parameter is
\begin{equation}
    \frac{\partial E_\theta}{\partial \theta_k}
    = 2\,\operatorname{Re}
      \frac{\langle \partial_k \Psi_\theta | H | \Psi_\theta \rangle}
           {\langle \Psi_\theta | \Psi_\theta \rangle}
    - 2\,\operatorname{Re}
      \frac{\langle \partial_k \Psi_\theta | \Psi_\theta \rangle}
           {\langle \Psi_\theta | \Psi_\theta \rangle}
      \cdot E_\theta,
    \label{eq:grad_raw}
\end{equation}
where $\partial_k \equiv \partial / \partial \theta_k$.
Introducing the logarithmic derivative
\begin{equation}
    O_k(x) = \frac{\partial \ln \Psi_\theta(x)}{\partial \theta_k}
                    = \frac{1}{\Psi_\theta(x)}
                      \frac{\partial \Psi_\theta(x)}{\partial \theta_k},
    \label{eq:log_deriv}
\end{equation}
and the local energy
\begin{equation}
    E_L(x) = \frac{H \Psi_\theta(x)}{\Psi_\theta(x)},
    \label{eq:local_energy}
\end{equation}
Eq.~\eqref{eq:grad_raw} can be recast into the compact covariance form
\begin{equation}
    \frac{\partial E_\theta}{\partial \theta_k}
    = 2\,\operatorname{Re} \Bigl\{
        \mathbb{E}_{x \sim |\Psi_\theta|^2}
        \bigl[ (E_L(x) - \mathbb{E}[E_L])\,
                (O_k(x) - \mathbb{E}[O_k])^* \bigr]
      \Bigr\}.
    \label{eq:force_formula}
\end{equation}
This expresses the energy gradient as the correlation function between the local energy and the logarithmic derivative, which can be estimated directly within Monte Carlo sampling~\cite{sorella2005}.

\subsection{Quantum Geometric Tensor and Natural Gradient}

Ordinary gradient descent $\delta\theta = -\eta \nabla_\theta E$ neglects the geometric structure of the parameter space: the distance between wave functions is not measured by the Euclidean metric $||\delta\theta||^2$, but by the Fubini-Study metric in the quantum state space.
For an infinitesimal parameter variation $\delta\theta$, the distance between normalized wave functions is
\begin{equation}
    d^2\bigl(|\Psi_{\theta+\delta\theta}\rangle,\, |\Psi_\theta\rangle\bigr)
    = \sum_{k,l} \mathcal{M}_{kl}\, \delta\theta_k \, \delta\theta_l + \mathcal{O}(\|\delta\theta\|^3),
    \label{eq:fs_metric}
\end{equation}
where $\mathcal{M}$ denotes the real part of the quantum geometric tensor (QGT),
namely the Fubini--Study metric used for real variational parameters,
and is defined as
\begin{equation}
    \mathcal{M}_{kl} = \operatorname{Re} \Bigl\{
        \frac{\langle \partial_k \Psi_\theta | \partial_l \Psi_\theta \rangle}
             {\langle \Psi_\theta | \Psi_\theta \rangle}
      - \frac{\langle \partial_k \Psi_\theta | \Psi_\theta \rangle}
             {\langle \Psi_\theta | \Psi_\theta \rangle}
        \frac{\langle \Psi_\theta | \partial_l \Psi_\theta \rangle}
             {\langle \Psi_\theta | \Psi_\theta \rangle}
      \Bigr\}.
    \label{eq:qgt_def}
\end{equation}
In terms of logarithmic derivatives, $\mathcal{M}_{kl}$ takes a succinct covariance form
\begin{equation}
    \mathcal{M}_{kl} = \operatorname{Re}\Bigl\{
        \mathbb{E}\bigl[O_k^* O_l\bigr]
      - \mathbb{E}\bigl[O_k^*\bigr]\,\mathbb{E}\bigl[O_l\bigr]
      \Bigr\}.
    \label{eq:qgt_cov}
\end{equation}
Natural gradient descent corrects the ordinary gradient by the metric $\mathcal{M}^{-1}$, with the parameter update
\begin{equation}
    \delta\theta = -\eta\, \mathcal{M}^{-1}\, \nabla_\theta E,
    \label{eq:nat_grad}
\end{equation}
which is the core formula of the Stochastic Reconfiguration (SR) method~\cite{sorella2005}.

\subsection{MCMC Estimation}

All expectation values $\mathbb{E}[\cdot]$ are taken with respect to the probability distribution
$p_\theta(x) = |\Psi_\theta(x)|^2 / \int dx'\,|\Psi_\theta(x')|^2$.
We use the Metropolis-Hastings algorithm to generate $N_s$ samples $\{x^{(i)}\}_{i=1}^{N_s}$ drawn from $p_\theta$, and replace population expectations by sample means:
\begin{equation}
    \mathbb{E}[f(x)] \approx
    \frac{1}{N_s} \sum_{i=1}^{N_s} f(x^{(i)}).
    \label{eq:mc_est}
\end{equation}
Due to autocorrelation among Markov chain Monte Carlo (MCMC) samples, in practice we use independent chains after thermalization and sample at intervals of several steps to reduce autocorrelation.
The Monte Carlo estimators for the gradient vector and the real QGT metric are
\begin{align}
    \hat{F}_k &= \frac{2}{N_s} \sum_{i=1}^{N_s}
        \operatorname{Re}\Bigl\{
            \bigl(E_L(x^{(i)}) - \bar{E}_L\bigr)\,
            \bigl(O_k(x^{(i)}) - \bar{O}_k\bigr)^*
        \Bigr\},
    \label{eq:grad_est} \\[4pt]
    \hat{\mathcal{M}}_{kl} &= \frac{1}{N_s} \sum_{i=1}^{N_s}
        \operatorname{Re}\Bigl\{
            O_k^*(x^{(i)})\, O_l(x^{(i)})
        \Bigr\}
      - \operatorname{Re}\bigl\{\bar{O}_k^*\, \bar{O}_l\bigr\},
    \label{eq:qgt_est}
\end{align}
where $\bar{E}_L$ and $\bar{O}_k$ are the sample means of the corresponding quantities.
For complex wave functions, $O_k(x)$ is generally complex. The real part used in
$\hat{\mathcal{M}}$ is positive semidefinite; the regularization term makes the
matrix positive definite when $\lambda_{\text{reg}}>0$.

\subsection{Regularization and Numerical Solution}

Directly inverting $\hat{\mathcal{M}}$ is numerically unstable for two reasons:
(1) the metric matrix has many near-zero eigenvalues, corresponding to nearly flat directions on the variational manifold;
(2) Monte Carlo sampling noise introduces statistical errors into the estimate of $\hat{\mathcal{M}}$.
We therefore employ a regularized inverse for the actual update:
\begin{equation}
    \delta\theta = -\eta\,
        (\hat{\mathcal{M}} + \lambda_{\text{reg}} 1)^{-1}\,
        \hat{F},
    \label{eq:sr_reg}
\end{equation}
where $\lambda_{\text{reg}}$ is an adaptive regularization parameter.
As $\lambda_{\text{reg}} \to \infty$, SR reduces to ordinary SGD; as $\lambda_{\text{reg}} \to 0$, pure natural gradient is recovered.
Since the parameter space dimension $N_p$ can reach tens to hundreds of thousands, explicitly constructing and inverting the $N_p \times N_p$ matrix $\hat{\mathcal{M}}$ is prohibitively expensive.
The actual implementation uses an iterative linear solver (conjugate gradient or MINRES), exploiting the matrix-vector product form of $\hat{\mathcal{M}}$ to avoid explicit matrix construction.
The numerical implementation is based on the JAX and NetKet frameworks~\cite{netket}.

\section{Scattering States}
\label{sec:appendix_scattering}

\begin{figure*}[htbp]
\centering
\includegraphics[width=\textwidth]{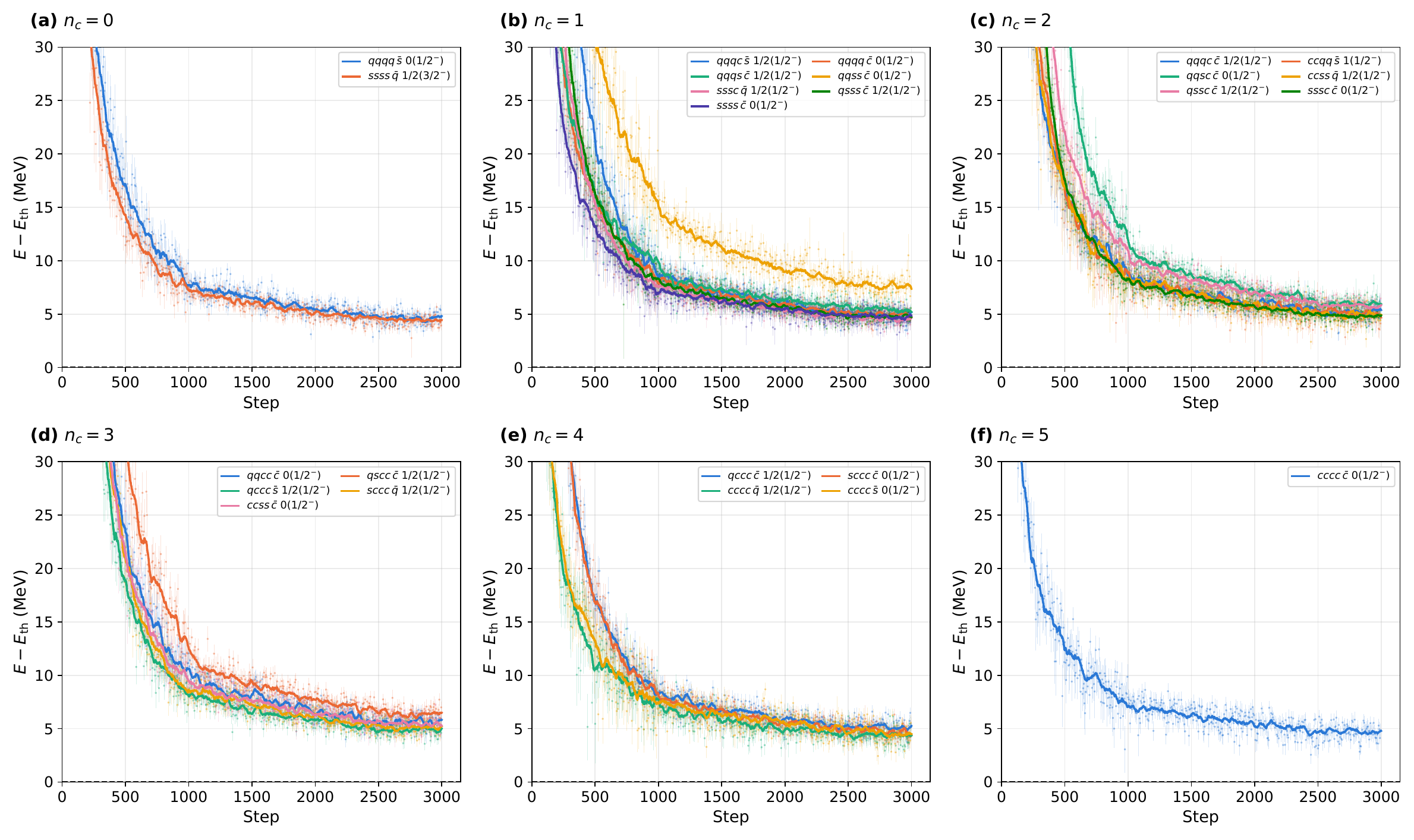}
\caption{
    Variational energy convergence of the S-wave scattering states in the lowest spin and isospin channels of the charmed pentaquark systems. The panels (a)--(f) correspond to the
    systems with different number of charm (anti)quarks ($n_c=0,\dots,5$). The labels
    indicate flavor content and quantum number $I(S^P)$. All energies are plotted
    with respect to the corresponding threshold. 
}
\label{fig:scattering_states}
\end{figure*}

\begin{figure*}[htbp]
\centering
\includegraphics[width=\textwidth]{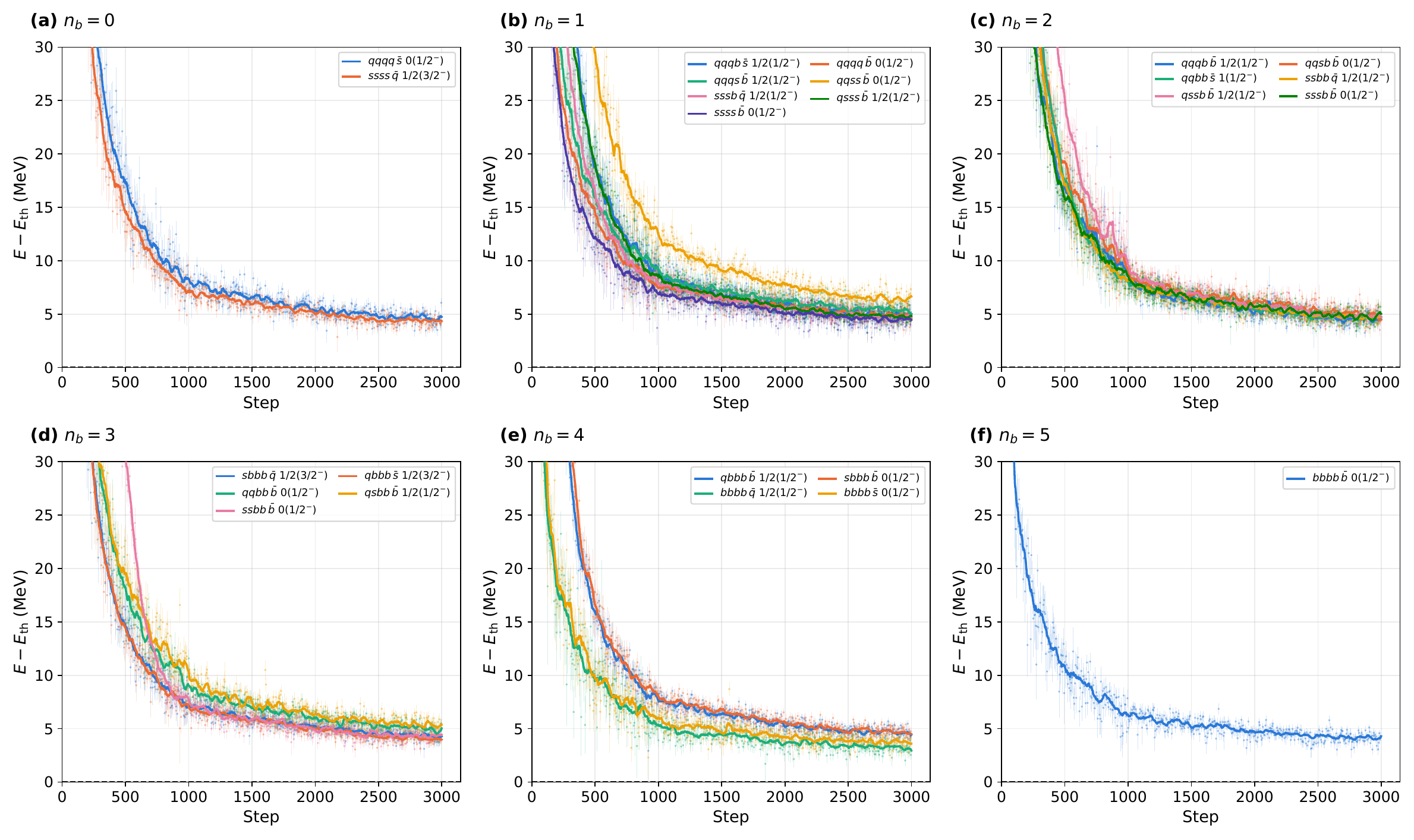}
\caption{
    Same as Fig.~\ref{fig:scattering_states} but for bottomed pentaquark systems.
}
\label{fig:scattering_states_b}
\end{figure*}

Figs.~\ref{fig:scattering_states} and
\ref{fig:scattering_states_b} show the energy convergence patterns of the S-wave scattering states in the lowest spin and isospin channels of the charmed and bottomed systems, respectively. The corresponding baryon--meson thresholds are listed in Table~\ref{tab:scattering_states}. The systems are grouped by the total number $n$ of heavy quarks and
heavy antiquarks, with $n=0,\dots,5$. Systems with higher spin and isospin, as well as those containing both charm and bottom quarks, are also calculated. Apart from the bound configurations in Table~\ref{tab:bound_states}, these systems exhibit similar convergence behavior.    All the converged energies approach the lowest thresholds from above.
The residual relative kinetic energy of the localized ansatz is around a few MeV. These scattering results provide the
baseline for identification of bound states. 

\begin{table}[htbp]
\centering
\caption{
    The lowest baryon--meson thresholds of the pentaquark systems, grouped by the total number $n$ of
    heavy quarks and heavy antiquarks.
}
\label{tab:scattering_states}
{
\setlength{\tabcolsep}{3pt}
\renewcommand{\arraystretch}{1.0}
\begin{tabular}{@{}clcc@{\hspace{0.7em}}lcc@{}}
\hline\hline
$n$ & System & $I(S^P)$ & Threshold & System & $I(S^P)$ & Threshold \\
\hline
0 & $qqqq\sbar$ & $0(1/2^-)$ & $ K N$
  & $ssss\qbar$ & $1/2(3/2^-)$ & $\bar K\Omega$ \\
\cline{1-7}
1 & $qqqc\sbar$ & $1/2(1/2^-)$ & $K\Lambda_c$
  & $qqqb\sbar$ & $1/2(1/2^-)$ & $K\Lambda_b$ \\
 & $qqqq\cbar$ & $0(1/2^-)$ & $\bar D N$
  & $qqqq\bbar$ & $0(1/2^-)$ & $ B N$ \\
 & $qqqs\cbar$ & $1/2(1/2^-)$ & $\bar D_sN$
  & $qqqs\bbar$ & $1/2(1/2^-)$ & $ B_sN$ \\
 & $qqss\cbar$ & $0(1/2^-)$ & $\bar D_s\Lambda$
  & $qqss\bbar$ & $0(1/2^-)$ & $ B_s\Lambda$ \\
 & $sssc\qbar$ & $1/2(1/2^-)$ & $\bar K\Omega_c$
  & $sssb\qbar$ & $1/2(1/2^-)$ & $\bar K\Omega_b$ \\
 & $qsss\cbar$ & $1/2(1/2^-)$ & $\bar D_s\Xi$
  & $qsss\bbar$ & $1/2(1/2^-)$ & $B_s\Xi$ \\
& $ssss\cbar$ & $0(1/2^-)$ & $\bar D_s^*\Omega$
  & $ssss\bbar$ & $0(1/2^-)$ & $B_s^*\Omega$ \\
\cline{1-7}
2 & $qqqc\cbar$ & $1/2(1/2^-)$ & $\eta_cN$
  & $qqqb\bbar$ & $1/2(1/2^-)$ & $\eta_bN$ \\
 & $ccqq\sbar$ & $1(1/2^-)$ & $ K\Xi_{cc}$
  & $qqbb\sbar$ & $1(1/2^-)$ & $ K\Xi_{bb}$ \\
 & $qqsc\cbar$ & $0(1/2^-)$ & $\eta_c\Lambda$
  & $qqsb\bbar$ & $0(1/2^-)$ & $\eta_b\Lambda$ \\
 & $ccss\qbar$ & $1/2(1/2^-)$ & $\bar K\Omega_{cc}$
  & $ssbb\qbar$ & $1/2(1/2^-)$ & $\bar K\Omega_{bb}$ \\
 & $qssc\cbar$ & $1/2(1/2^-)$ & $\eta_c\Xi$
  & $qssb\bbar$ & $1/2(1/2^-)$ & $\eta_b\Xi$ \\
 & $sssc\cbar$ & $0(1/2^-)$ & $\bar D_s\Omega_c$
  & $sssb\bbar$ & $0(1/2^-)$ & $\Upsilon\Omega$ \\
\cline{1-7}
3 & $qqcc\cbar$ & $0(1/2^-)$ & $\eta_c\Lambda_c$
  & $qqbb\bbar$ & $0(1/2^-)$ & $\eta_b\Lambda_b$ \\
 & $qscc\cbar$ & $1/2(1/2^-)$ & $\eta_c\Xi_c$
  & $qsbb\bbar$ & $1/2(1/2^-)$ & $\eta_b\Xi_b$ \\
 & $qccc\sbar$ & $1/2(1/2^-)$ & $D_s\Xi_{cc}$
  & $qbbb\sbar$ & $1/2(3/2^-)$ & $K\Omega_{bbb}$ \\
 & $sccc\qbar$ & $1/2(1/2^-)$ & $D\Omega_{cc}$
  & $sbbb\qbar$ & $1/2(3/2^-)$ & $\bar K\Omega_{bbb}$ \\
 & $ccss\cbar$ & $0(1/2^-)$ & $\bar D_s\Omega_{cc}$
  & $ssbb\bbar$ & $0(1/2^-)$ & $\eta_b\Omega_b$ \\
\cline{1-7}
4 & $qccc\cbar$ & $1/2(1/2^-)$ & $\eta_c\Xi_{cc}$
  & $qbbb\bbar$ & $1/2(1/2^-)$ & $\eta_b\Xi_{bb}$ \\
 & $sccc\cbar$ & $0(1/2^-)$ & $\eta_c\Omega_{cc}$
  & $sbbb\bbar$ & $0(1/2^-)$ & $\eta_b\Omega_{bb}$ \\
 & $cccc\qbar$ & $1/2(1/2^-)$ & $D^*\Omega_{ccc}$
  & $bbbb\qbar$ & $1/2(1/2^-)$ & $\bar B^*\Omega_{bbb}$ \\
 & $cccc\sbar$ & $0(1/2^-)$ & $ D_s^*\Omega_{ccc}$
  & $bbbb\sbar$ & $0(1/2^-)$ & $\bar B_s^*\Omega_{bbb}$ \\
\cline{1-7}
5 & $cccc\cbar$ & $0(1/2^-)$ & $J/\psi\Omega_{ccc}$
  & $bbbb\bbar$ & $0(1/2^-)$ & $\Upsilon\Omega_{bbb}$ \\
\hline\hline
\end{tabular}}
\end{table}

\bibliography{NN-VMC_Pentaquark}

\end{document}